\documentclass[referee]{raa}
\usepackage{graphicx,times}
\usepackage{natbib}
\usepackage{amssymb,amsmath}
\usepackage[pagebackref=true]{hyperref}

\begin{document}

   \title{Direct Imaging for the Debris Disk around $\epsilon$~Eridani with the Cool-Planet Imaging Coronagraph}

 \volnopage{ {\bf 20XX} Vol.\ {\bf X} No. {\bf XX}, 000--000}
   \setcounter{page}{1}

   \author{Chun-Hui Bao\inst{1,2}, Jiang-Hui Ji \inst{1,2,3}, Gang Zhao\inst{4,5}, Yi-Ming Zhu\inst{4,5}, Jiang-Pei Dou \inst{4,5}, Su Wang\inst{1,3}, Yao Dong \inst{1,3}
   }

   \institute{
CAS Key Laboratory of Planetary Sciences, Purple Mountain Observatory, Chinese Academy of Sciences, Nanjing 210023, China;{\it jijh@pmo.ac.cn}\\
\and
School of Astronomy and Space Science, University of Science and Technology of China, Hefei 230026, China\\
\and
CAS Center for Excellence in Comparative Planetology, Hefei 230026, China\\
\and
CAS Key Laboratory of Astronomical Optics \& Technology, Nanjing Institute of Astronomical  Optics \& Technology, Nanjing 210042, China
\and
Nanjing Institute of Astronomical Optics \& Technology, Chinese Academy of Sciences, Nanjing 210042, China
\vs \no \\
   {\small Received 20XX Month Day; accepted 20XX Month Day}
}
\abstract{We analyze the inner debris disk around $\epsilon$~Eridani using simulated observations with the Cool-Planet Imaging Coronagraph (CPI-C). Using the radiative transfer code \texttt{MCFOST}, we generate synthetic scattered-light images and spectral energy distributions  for three disk models that differ in inclination and radial extent, and compare these results with the anticipated performance of CPI-C. CPI-C can resolve disk structures down to $\sim$3~au, offering substantially finer spatial resolution than existing HST/STIS and Spitzer/IRS observations. Recovered inclinations and radial extents closely match the input models, constraining the disk geometry and informing potential planet-disk interactions in the $\epsilon$~Eri system. Although the cold Jupiter-like planet $\epsilon$~Eri~b is not detected in our simulations, polarimetric methods may enable detection of its reflected light. These results highlight the capability of next-generation coronagraphs to probe cold dust in nearby planetary systems.
}
   \authorrunning{C.-H. Bao et al.}
   \titlerunning{Imaging $\epsilon$~Eridani debris disk with CPI-C}
   \maketitle

\section{Introduction}
\label{sect:intro}
Since the first detection of a debris disk around Vega by the IRAS satellite, more than one thousand stars have been identified to host debris disks, primarily through infrared excess or direct imaging. However, due to technological limitations, only about 10\% of these disks have been spatially resolved, and only a small fraction of the host stars are known to harbor planets \citep{Cao2023}. Debris disks are optically thin dusty structures thought to originate from the collisional grinding of planetesimals, offering critical insights into the architecture and evolution of planetary systems \citep{Liu2020}. Their morphology such as rings, gaps, spirals, or asymmetries can be sculpted by planet-disk interactions, including gravitational perturbations or resonant trapping \citep{Jin2016,Dong2017,Bae2018, Jin2019, Huang2020, Cimerman2021, Paardekooper2023, Cimerman2024, Huang2025}. Both theoretical and numerical studies have revealed that pressure bumps and instabilities in protoplanetary disks can lead to the formation of long-lived substructures, including vortices and spiral arms, which may persist into the debris disk phase \citep{Huang2018,Huang2019}. These structures are not only critical to understanding disk evolution, but also serve as indirect tracers of unseen planetary companions.

Located at a distance of 3.2 pc, Epsilon Eridani (hereafter $\epsilon$ Eri) is the nearest known star with a confirmed debris disk. It is also considered a young solar analog, with an estimated age of 400 - 800 Myr. Its fundamental parameters include an effective temperature of approximately 5020 K \citep{Rosenthal2021}, a radius of 0.738 $R_\odot$ \citep{Rains2020}, and a mass of 0.88 $M_\odot$ \citep{Baines2012}.

Because most stars are located more than ten parsecs away, the inner working angle (IWA), typically larger than 0.3 arcseconds, limits the ability to detect warm dust close to the host star. Moreover, previous studies have largely focused on the infrared and submillimeter wavelengths, where emission from cold dust is more prominent. Therefore, as a nearby system, $\epsilon$ Eri offers a valuable opportunity to resolve its inner debris disk. {The system hosts a complex, multi-belt architectures similar to that of the solar system, including a broad outer ring centered near 60 au \citep{Greaves1998}, a middle belt near 20 au \citep{Backman2009}, and an innermost warm belt inferred at 3 au \citep{Su2017}. This layered structure, together with the presence of the cold Jupiter $\epsilon$ Eri b orbiting at 3.5 au, makes $\epsilon$ Eri an ideal target for CPI-C to detect the inner debris material and planetary companions, as well as their interactions.}
 
This hierarchical debris disk structure, has been observed at infrared and submillimeter wavelengths. \citet{Aumann1985} first inferred the existence of the disk through infrared excess detected by the IRAS satellite. A Kuiper Belt analog, peaking approximately at 60 au, was resolved in the submillimeter, with a derived inclination of $\sim 25^\circ$ \citep{Greaves1998, Greaves2005}. Using the imaging and spectroscopic data from \textit{Spitzer} and the Caltech Submillimeter Observatory, \citet{Backman2009} modeled the outer belt as a combination of large ice grains and slightly small silicate grains, distributed between 35 and 90 au. \textit{Spitzer} observations also reveal the presence of two inner belts. Based on the model of \citet{Backman2009}, one belt composed of silicate grains was inferred to lie near 20 au. An additional innermost belt is inferred to be located at 3 au, though its detailed composition and structure remain unconstrained due to the limited resolution of \textit{Spitzer}.

\citet{Greaves2014} provided further constraints on the middle and outer belt using \textit{Herschel} imaging, identifying structures spanning 12-16 and 54-68 au, respectively. These findings are broadly consistent with the results of \citet{MacGregor2015}, who placed the center of the outer belt at $64.4^{+2.4}_{-3.0}$ au, with a FWHM of $20.2^{+6.0}_{-8.2}$ based on Submillimeter Array imaging observations. More recent ALMA observations refined the structure of the outermost belt, suggesting a narrower ring centred at 69 au, with a width of 11 $\sim$ 13 au, and an inclination of about $34^\circ$ \citep{Booth2017}. Additionally, clumpy substructures detected in ALMA images suggest the possible presence of another planet at around 40 au, potentially in the process of migration \citep{Booth2023}. These features suggest possible dynamical sculpting by planetary companions and highlight $\epsilon$ Eri as an ideal testbed for studying planet-disk interactions across multiple wavelengths.

Although the outer two belts of $\epsilon$ Eri have been well resolved, the innermost disk remains poorly understood. Nulling interferometry observations at 8-13 $\mu\mathrm{m}$ have revealed IR excess emission from the inner region of $\epsilon$ Eri \citep{Ertel2020}. The Hubble Space Telescope's Space Telescope Imaging Spectrograph (HST/STIS) provided the first detection of the debris system in the visible band \citep{Wolff2023}. However, the inner belt was not detected because its surface brightness is at least two orders of magnitude below the STIS sensitivity threshold \citep{Wolff2023}.

Using the SPHERE and ZIMPOL imaging polarimeter, \citet{Tschudi2024} observed the inner region of $\epsilon$ Eri at visible wavelengths, but did not achieve a significant detection of warm dust due to a contrast limit of 15 $\mathrm{mag} \ \mathrm{arcsec}^{-2}$. Observations with the HST/STIS WEDGE A coronagraph in the optical band also failed to detect the inner disk, placing an upper brightness limit of 6~$\mathrm{mJy}$ at $0\farcs6$ \citep{Sai2024}.

$\epsilon$ Eri is known to host a cold giant planet, $\epsilon$~Eri~b, located at approximately 3.5~au. Its orbital parameters have been derived from radial velocity and astrometric observations \citep{Mawet2019, Llop2021, Thompson2025}, but significant discrepancies remain, particularly in the inclination and planetary mass. Even two or three direct imaging observations can substantially improve the precision of the planet's orbital solution \citep{Bruna2023}. Moreover, combining direct imaging with radial velocity and astrometry can place tighter constraints on the planetary mass.

The $\epsilon$ Eri system is also a priority target for several upcoming space missions. {The James Webb Space Telescope (JWST) has already observed $\epsilon$ Eri through several programs, using both the Mid-Infrared Instrument (MIRI) and the Near-Infrared Camera (NIRCam), and additional proposals are still ongoing in later cycles. Recently, \citet{Llop-Sayson:2025:arXiv} reported their results, placing new contrast limits beyond 1". A point source is found within $1-\sigma$ at the expected position of planet by \citet{Thompson2025}. However, it is not statistically significant due to the proximity of a bright hexpeckle. Moreover, the observed surface brightness distribution is consistent with the expected inclination of the outer disk by \citet{Booth2017}. However, due to the inner working angle of the JWST coronagraph, the regions within 1" are severely contaminated, and consequently, no signal from the innermost belt is detected.} In addition, the Coronagraph Instrument on the Nancy Grace Roman Space Telescope (Roman), scheduled for launch in 2027, is designed to detect scattered light from the unresolved inner disk \citep{Poberezhskiy2020}.

The CPI-C, one of the instruments onboard the China Space Station Telescope (CSST) \citep{CSST2025}, is designed to detect Jovian planets and debris disks with contrasts reaching $10^{-8}$ (Zhao et al., 2025; Zhu et al., 2025). In this work, we assess the feasibility of imaging the innermost disk and potential planets in the $\epsilon$~Eri system using CPI-C. We simulate high-contrast direct imaging observations by modeling both the inner debris belt and the long-period giant planet $\epsilon$~Eri ~b, evaluating CPI-C's ability to spatially resolve complex circumstellar structures. Our results show that CPI-C can distinguish extended disk features from stellar speckles. We also explore the use of polarimetric imaging to enhance detectability, although its effectiveness is limited by systematic noise and long integration times. These findings demonstrate CPI-C's potential to probe planet-disk interactions and to constrain dynamical architectures in nearby planetary systems.

The remainder of this paper is structured as follows. Section~\ref{sect:CPI-C} provides a brief overview of the CPI-C instrument. Section~\ref{sect:methods}  describes the modeling methods employed for the debris disk and planet. Section~\ref{sect:processing}, we present the data processing pipeline and analyze the imaging results, including the recovered disk parameters and CPI-C's detection limits. Finally, Section~\ref{sect:Discussion} summarizes the key findings and discusses their broader implications.

\section{CPI-C and Observations}
\label{sect:CPI-C}
We provide a brief overview of CPI-C, which is described in detail by Zhu et al.~(2025) and Zhao et al.~(2025). Utilizing a combination of deformable mirrors and focal plane masks, CPI-C generates two square-shaped dark zones on one side of the PSF core, capable of achieving a contrast ratio of $10^{-8}$. The instrument is specifically designed to directly detect and characterize cold exoplanets and circumstellar debris disks, with particular emphasis on planetary systems analogous to the Solar System. Its primary science goals include the detection of cold Jupiter-like planets at wide orbital separations (typically beyond 5~au), which are crucial for understanding the architectures and formation mechanisms of planetary systems.

In addition, CPI-C aims to spatially resolve faint scattered light from inner debris belts. These observations can constrain dust grain properties, spatial distribution, and dynamical structures potentially shaped by unseen planets. By integrating coronagraphy with polarimetric and multi-band imaging, CPI-C also seeks to characterize planetary atmospheres through reflected-light spectra probing their composition, cloud structure, and albedo. These capabilities will enable CPI-C to significantly enhance our understanding of planet-disk interactions and the diversity of planetary systems in the solar neighborhood.

The dark zones span from 4 to 16 $\lambda/D$, corresponding to angular separations of approximately 240 to 990 milli-arcsecond at 600 nm. For the target $\epsilon$ Eri, this angular range translates to projected separations of 0.77 to 3.2~au on the sky. This region encompasses both the innermost debris disk and the candidate planet of interest, as discussed in the following section. CPI-C covers a broad wavelength range from the optical to near-infrared, employing eight narrow band filters centered at 565, 661, 743, 883, 940, 1265, 1425, and 1542nm. The basic parameters of CPI-C are listed in Table \ref{tab:CPIC}. Compared to existing high-contrast imaging facilities, such as the James Webb Space Telescope (JWST), VLT/SPHERE, and the upcoming Roman Space Telescope, CPI-C is specifically optimized for detecting faint, cold exoplanets and inner debris structures at small angular separations. While JWST provides exceptional sensitivity in the mid-infrared, its coronagraphic inner working angle (IWA) restricts access to regions within a few astronomical units around nearby stars. VLT/SPHERE offers high spatial resolution and adaptive optics correction in the near-infrared, but is limited by atmospheric turbulence and reduced contrast performance at small IWAs. Roman will feature advanced coronagraphy in space; however, its broad-band optical coverage may not be optimal for imaging scattered light from cold dust. For comparison, key parameters of Roman are also listed Table \ref{tab:CPIC} \citep{Bailey2023}.

\begin{table}
\bc
\begin{minipage}[]{120mm}
\caption[]{Basic parameters of CPI-C and Roman.\label{tab:CPIC}}
\end{minipage}
\setlength{\tabcolsep}{1pt}
\small
 \begin{tabular}{cccc}
  \hline\noalign{\smallskip}
Parameters& CPI-C & & Roman \\
  \hline\noalign{\smallskip}
 Inner Working Angle (IWA) & $4\lambda/D$ && $3\lambda/D$\\
 Outer Working Angle (OWA) & $16\lambda/D$ & &$9\lambda/D$\\
Wavelength & 500 - 1600 nm & & 500 - 900 nm\\
Aperture (m) & 2 \textrm{m} & &2.4 \textrm{m}\\
Dark Zone Contrast &  $ 1 \times 10^{-8}$&& $1 \times 10^{-8}$ \\
\noalign{\smallskip}\hline
\end{tabular}
\ec
\end{table}

\section{Generating disk and planet models}
\label{sect:methods}
\subsection{Disk Intensity Modeling}

To simulate observation of $\epsilon$ Eri, we model the intensity distribution of the debris disk using the radiative transfer modeling software \texttt{MCFOST} \citep{Pinte2006, Pinte2009}. As introduced in the previous section, the system includes three dust belts. Given the limited knowledge about the innermost belt, we construct three representative models to evaluate CPI-C's detection capabilities. The key parameters used in the models are listed in Table \ref{tab:mcfost}. Although the detailed parameters of the outer two belts differ among previous studies, we adopt a unified model for them, because their emission predominantly contributes to the far-infrared and sub-millimeter bands and has little influence on the CPI-C wavelength coverage.

In the first model, the innermost belt is assumed to be nearly face-on. Inclination plays a critical role in high-contrast imaging. We initially assume the innermost belt shares the same inclination as the outer two belts, $34^\circ$, provided by \citet{Booth2017}. The radial extent of this belt is informed by previous analyses \citep{Backman2009, Su2017}, which suggest that it lies within 3 au. Following the methodology of \citet{Su2017}, we further constrain the belt's location by considering the planet-induced chaotic zones \citep{Mustill2012,Morrison2015}. Assume an eccentricity less than 0.1, and a mass less than 1 $M_\mathrm{J}$ \citep{Mawet2019,Thompson2025}, we estimate the innermost belt spans from 1.5 to 2.5 au.

However, recent studies suggest a significant misalignment of $42.7^{\circ}{}^{+11}_{-9.5}$  between the stellar spin axis and the outer disk plane \citep{Hurt2023}. Therefore, in the second model, we adopt an alternative configuration in which the innermost belt is aligned with the stellar equator. This corresponds to an inclination of $70^\circ$, as reported by \citet{Giguere2016}.

Moreover, motivated by the findings of \citet{Sai2024} and \citet{Ertel2020}, who suggest a broad dust region, we adopt a third model featuring a continuous disk that extends from 0.1 to 3 au. The remaining parameters, including dust mass and grain composition, are chosen such that the resulting surface brightness remains consistent with observational constraints. Specifically, the modeled brightness is limited to 6 $\mathrm{mJy}\ \mathrm{arcsec}^{-2}$ and  $\Delta \mathrm{mag} = 15 \ \mathrm{arcsec}^{-2}$, based on the detection threshold by \citet{Tschudi2024} and \citet{Sai2024}, respectively.
\begin{table}
\bc
\begin{minipage}[]{100mm}
\caption[]{Parameters of innermost belt used in \texttt{MCFOST}\label{tab:mcfost}}
\end{minipage}
\setlength{\tabcolsep}{1pt}
\small
 \begin{tabular}{cccccccc}
  \hline\noalign{\smallskip}
Name& Model A & Model B&  Model C\\
  \hline\noalign{\smallskip}
Location (au) & 1.5 - 2.5 & 1.5 - 2.5 & 0.1-3.0\\
Inclination ($^\circ$) & 34& 70& 34\\
Position angle ($^\circ$) & -4& -4& -4\\
Dust Mass ($M_\odot$) & $5.0\times 10^{-12}$ & $3.0\times 10^{-13}$ & $1.0\times 10^{-13}$\\
Component& 100\% astrosilicates & 100\% astrosilicates& 100\% astrosilicates \\
Minimum Grain size ($\mu$m) & 1 & 3& 0.1 \\
Maximum Grain size ($\mu$m) & 1000 & 1000& 1000 \\
  \noalign{\smallskip}\hline
\end{tabular}
\ec
\end{table}

\begin{figure}
   \centering
  \includegraphics[width=10cm, angle=0]{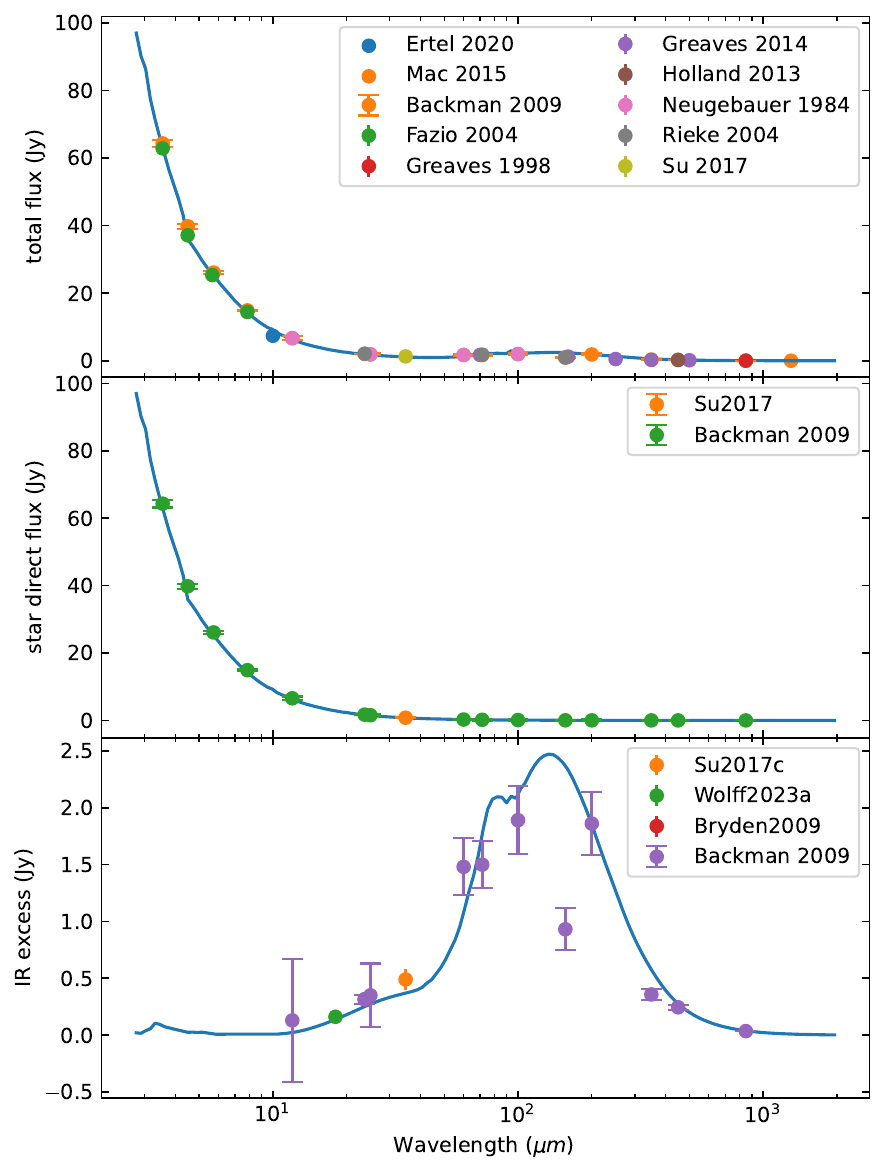}
   \caption{Spectral energy distribution (SED) of Model A, computed using \texttt{MCFOST}. From top to bottom: (1) total flux including both stellar and disk components; (2) stellar photosphere only; (3) infrared excess derived by subtracting the stellar model from the total emission. The x-axis shows wavelength in microns, and the y-axis shows flux density in Jy. Solid lines represent the model outputs, while data points from the literature (\citep{Neugebauer1984}, \citep{Greaves1998}, \citep{Fazio2004}, \citep{Rieke2004}, \citep{Backman2009}, \citep{Bryden2009}, \citep{Holland2013}, \citep{MacGregor2015}, \citep{Su2017}, \citep{Ertel2020}, \citep{Wolff2023}) are overplotted for comparison.}
   \label{fig:sed}
\end{figure}

\begin{figure}
   \centering
  \includegraphics[width=15cm, angle=0]{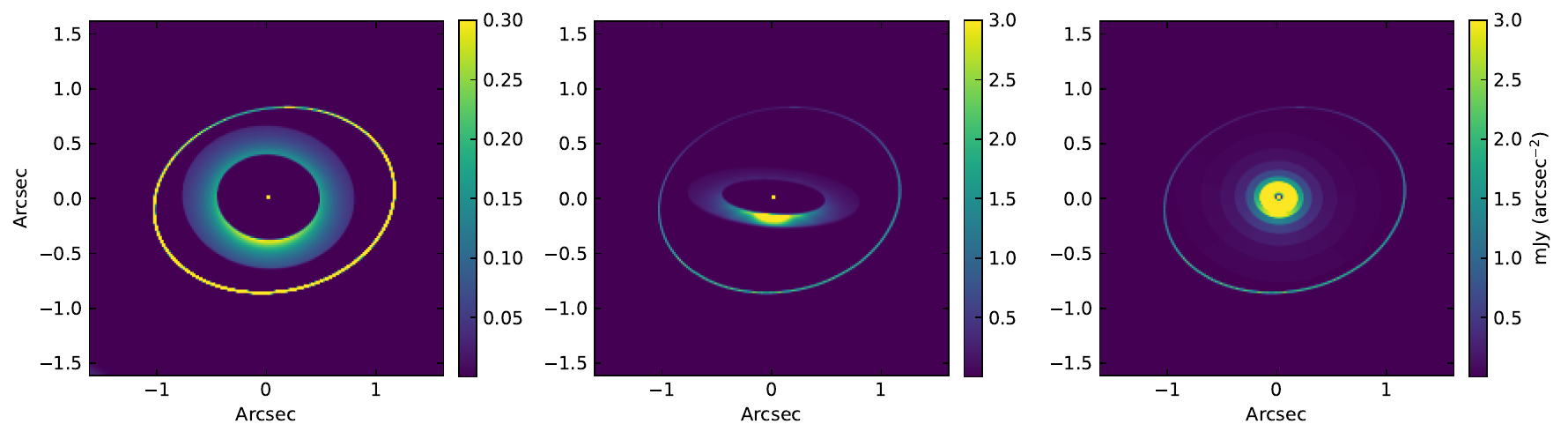}
   \caption{Local view of the simulated innermost disk flux distribution models of $\epsilon$ Eri. \textit{Left Panel}: Low inclination disk model. \textit{Middle Panel}: High inclination disk model. \textit{Right Panel}:Continuous disk model. The outer line represents planetary trace based on orbital parameters from \citep{Thompson2025}.}
   \label{fig:ideal}
\end{figure}

Figure \ref{fig:sed} presents the Model A spectral energy distribution (SED) computed with \texttt{MCFOST}, with panels showing the total emission, stellar photosphere, and infrared excess from the dust disk. The model outputs are overplotted as continuous curves, while photometric data points from the literature are shown for comparison. The IR excess match reasonably well with the observed IR excess at mid-IR wavelengths. As our disk model only focus on the innermost belt, the SED beyond 25 $\mu\mathrm{m}$ is not matched perfectly, which is from the outer ring. We find that the warm belt primarily contributes to the flux from the optical to mid-infrared wavelengths, where observational constraints are currently limited. Simulated disk images for all three models are presented in Figure~\ref{fig:ideal}, only a zoomed-in portion of the full field of view is shown. For comparison, the projected orbit of $\epsilon$~Eri b, based on parameters from \citet{Thompson2025}, is overplotted as a bright trace, which will be described in Section \ref{subsect:model_pl}. The left, middle, and right panels correspond to Models A, B, and C, respectively.

\subsection{Planetary contrast modeling} \label{subsect:model_pl}
\begin{figure}
   \centering
  \includegraphics[width=10cm, angle=0]{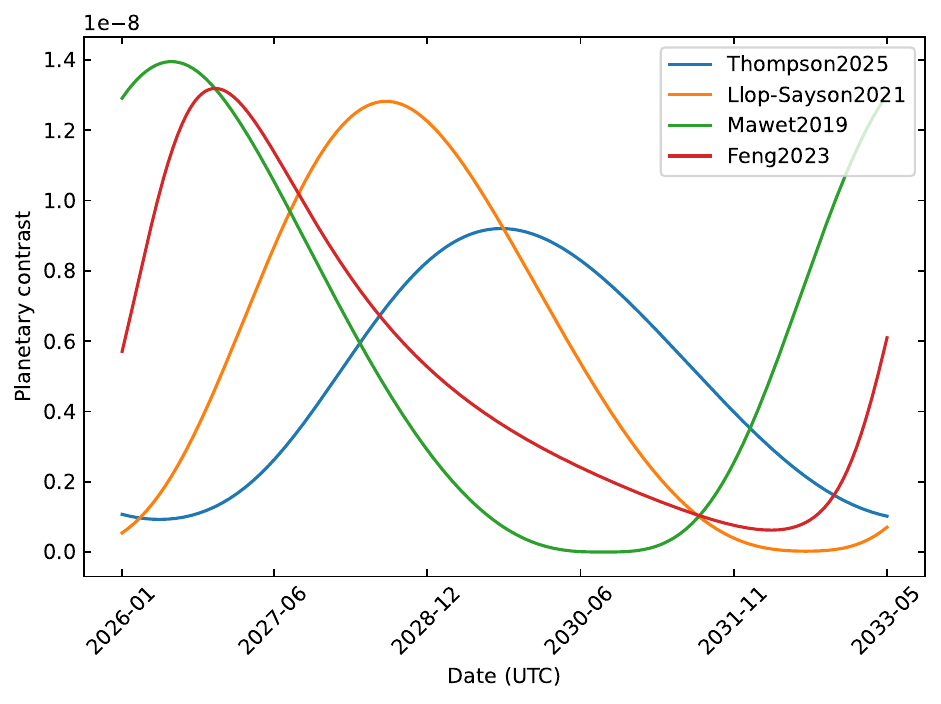}
   \caption{Estimated planetary contrast at visible band based on various orbital parameter sets. Modeled planetary contrast $C_\mathrm{p}$ of $\epsilon$ Eri b as a function of time.
   The blue, orange, green and red curves are based on orbital solutions adopted from \citet{Thompson2025}, \citet{Llop2021}, \citet{Mawet2019}, and \citet{Feng2023}, respectively. All curves assume a wavelength-independent geometric albedo of 0.5. The contrast peaks at values around $10^{-8}$, which is near the designed detection threshold of CPI-C.
}
   \label{fig:pl_phase}
\end{figure}

The Jovian planet $\epsilon$ Eri b constitutes another key component of the system, situated between the innermost and middle debris belts. As a cold Jupiter-like planet, its emission is dominated by reflected starlight rather than thermal radiation. The intensity contrast $C_\mathrm{p}$ between the planet and the star can be estimated as:
\begin{equation}\label{eq:contrast}
   C_\mathrm{p} = A_g(\lambda) \Phi (\beta) \frac{R_\mathrm{p}^2}{d^2} ,
\end{equation}
where $A_g$ is the geometric albedo dependent on wavelength $\lambda$, $\Phi (\beta)$ is the phase function of the phase angle $\beta$, $R_\mathrm{p}$ is the planetary radius, and $d$ is the instantaneous star-planet distance. The phase angle $\beta$ is defined as the angle between the planet-star and the planet-observer directions. It can be calculated from the Keplerian orbital elements as:
\begin{equation}
   \cos{\beta} = \sin{I} \sin{(\theta + \omega)},
\end{equation}
where $I$ is the orbital inclination, $\theta$ is the true anomaly, and $\omega$ is the argument of periapsis. The projected position of the planet on the sky plane, denoted by $x(t)$ and $y(t)$, is given by:
\begin{equation}\label{eq5}
\begin{aligned}
   x (t) &= A X(t) + F Y(t) \, , \\
   y (t) &= B X(t) + G Y(t) \, ,
   \end{aligned}
\end{equation}
where the parameters in these equations are derived from the Thiele-Innes equations \citep{Thiele1883}:
\begin{equation}\label{equ:TI}
   \begin{aligned}
   \left\{
       \begin{array}{ll}
       &A=  a (\cos \Omega \cos \omega - \sin \Omega \sin \omega \cos I)  \, , \\
       &B=  a (\sin \Omega \cos \omega + \cos \Omega \sin \omega \cos I)  \, , \\
       &F=  a (-\cos \Omega \sin \omega - \sin \Omega \cos \omega \cos I) \, , \\
       &G=  a (-\sin \Omega \sin \omega + \cos \Omega \cos \omega \cos I) \, , \\
       &X (t) = \cos E(t) - e  \, , \\
       &Y (t) = \sqrt{1-e^2} \sin E(t)  \, . \\
       \end{array}
   \right.
   \end{aligned}
\end{equation}
where $a$, $\Omega$, $e$, and $E$ are the planetary semi-major axis, longitude of the ascending node, eccentricity, and eccentric anomaly, respectively.
Thus, by combining the known orbital parameters with an assumed geometric albedo spectrum, we calculate the expected planetary over the coming years, as shown in Figure \ref{fig:pl_phase}. The four curves correspond to different sets of orbital parameters adopted from the literature, specifically \citet{Thompson2025}, \citet{Llop2021}, \citet{Mawet2019}, and \citet{Feng2023}. Each curve reflects the variation of contrast as a function of viewing geometry along the orbit. The simple Lambert reflectance phase function has been adopted. We find that he contrast peaks are all located around $\sim$10$^{-8}$, which is comparable to the expected detection limit of CPI-C.

\section{Data processing and analyses} \label{sect:processing}

\subsection{Simulated imaging of CPI-C}
\begin{figure}
   \centering
  \includegraphics[width=12cm, angle=0]{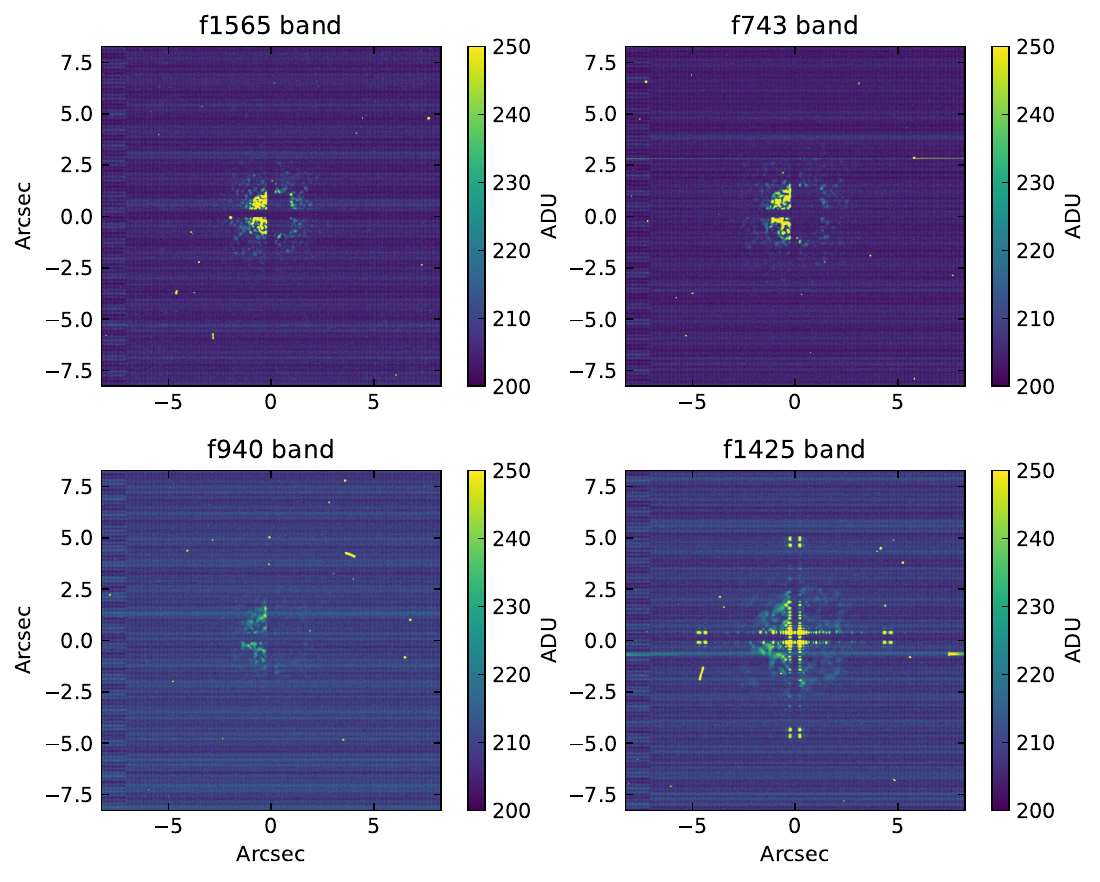}
   \caption{Simulated raw coronagraphic images of the $\epsilon$~Eri system in four CPI-C bands, generated with \texttt{CPISM}. Top left: F565; top right: F743; bottom left: F940; bottom right: F1425. Each image includes simulated detector effects, cosmic ray hits, and wavefront-induced speckles. Two square regions, slightly offset to the right of image center, denote the dark zones where starlight suppression is most effective.}
   \label{fig:raw_img}
\end{figure}

To simulate the imaging observations for $\epsilon$ Eri, we utilize the \texttt{CPISM} software\footnote{\url{https://csst-tb.bao.ac.cn/code/csst-sims/csst_cpic_sim}} developed by the CPI-C team (Zhao et al.,~2025; Zhu et al.,~2025). This software integrates a detailed model of the CPI-C optical system with a realistic end-to-end simulation of the imaging process, incorporating diffraction, speckle noise, and detector characteristics, based on the specifications of the designed EMCCD camera. For our simulations, we adopt an exposure time of 50 seconds per frame, which provides sufficient photons from scattered light while preventing detector saturation.  Based on \texttt{CPISM} thermal modeling, we verify that a total integration time of $50\ \mathrm{s} \times 20$ frames can be achieved while maintaining the camera temperature at approximately $-80^\circ\mathrm{C}$. Under these conditions, the cosmic ray rate and stellar core saturation level are both within acceptable limits, and any residual effects can be effectively mitigated through the post-processing.  The raw output images based on the Model A from \texttt{CPISM} simulations are shown in Figure \ref{fig:raw_img}, covering four representative CPI-C bands. Each image includes the dominant noise sources (bias, speckles, and cosmic rays) and illustrates the system's diffraction pattern. The central region, slightly offset to the right, contains two square dark zones where starlight is actively suppressed by the coronagraph; these zones are the primary regions of interest for planet and disk detection. Among the bands shown, the F565 band exhibits the highest image quality, exhibiting both minimal speckle residuals and optimal contrast within the dark zone.

The raw images generated by \texttt{CPISM} incorporate various instrumental and observational effects, including cosmic rays, detector bias, and dark current. To correct for these effects, we apply standard preprocessing with the \texttt{ccdproc} package \citep{CCDPROC}, producing dark-subtracted, bias-corrected, and flat-fielded frames suitable for scientific analysis.

\begin{figure}
   \centering
  \includegraphics[width=15cm, angle=0]{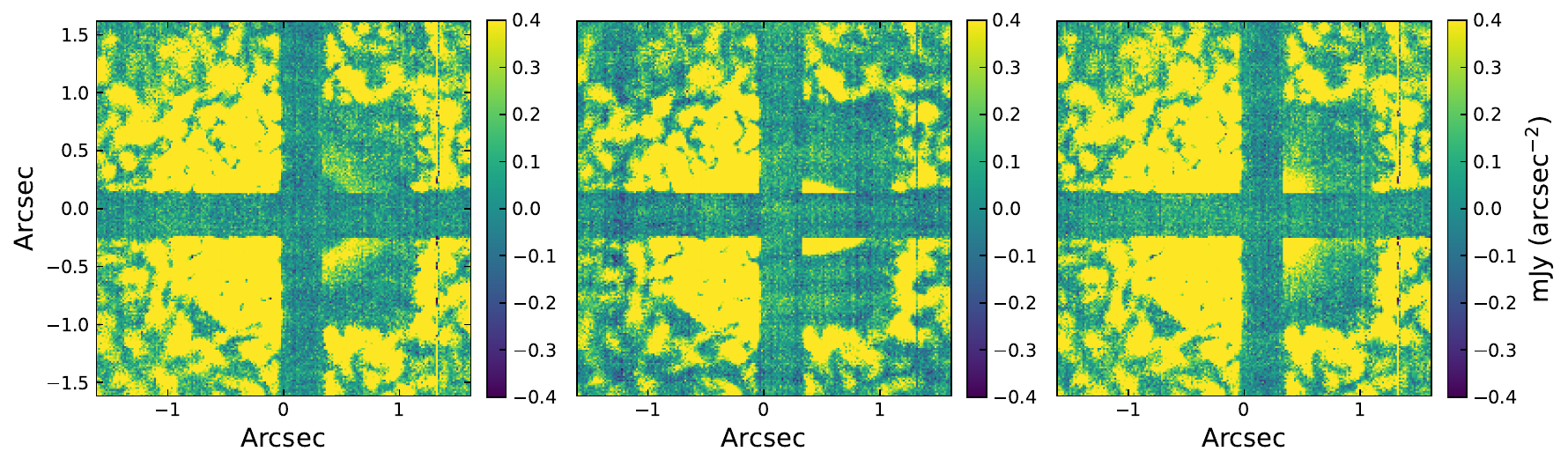}
   \caption{PSF-subtracted simulated images for Models A, B, and C (from left to right), each showing the central $\sim$3$\arcsec$ region in the $F565$ filter. All three models reveal disk structures within the dark zones. Model A shows a well-defined belt with clearly identifiable inner and outer edges. Model B, due to its high inclination, displays the most prominent and spatially extended features. Model C presents a continuous disk with a smooth brightness gradient, decreasing gradually from the center outward.}
   \label{fig:preprocessing}
\end{figure}

Following preprocessing, we perform classical PSF subtraction as part of the high-contrast data reduction pipeline. For this purpose, we simulate a reference star observed under identical instrumental and observational conditions, assuming the same apparent magnitude and spectral type as $\epsilon$~Eri to ensure an appropriate PSF match. {This reference frame was directly subtracted from the target images without applying more advanced algorithms such as principal component analysis (PCA) or KLIP \citep{Wang:2015:ascl}. While simple, this method provides a baseline estimate of CPI-C's subtraction performance and allows us to assess the disk detectability in idealized conditions.} The resulting PSF-subtracted images are shown in Figure~\ref{fig:preprocessing}, where significant improvements in contrast are evident, particularly within the dark zones. Figure~\ref{fig:preprocessing} presents the PSF-subtracted images for the three disk models (A, B, and C), with each panel displaying only the central $\sim$3~arcseconds corresponding to the innermost belt. Disk structures are clearly detected in all three models. Model~B exhibits the most distinct morphology, largely due to the bright forward-scattering peak associated with its high inclination. In Model~A, the belt appears as a relatively sharp structure with clearly distinguishable inner and outer edges, consistent with a narrow and confined dust distribution. By contrast, Model~C shows a more extended and continuous emission profile, with surface brightness gradually decreasing from the inner to outer regions, as expected for a broad-ring morphology.

\begin{figure}
   \centering
  \includegraphics[width=15cm, angle=0]{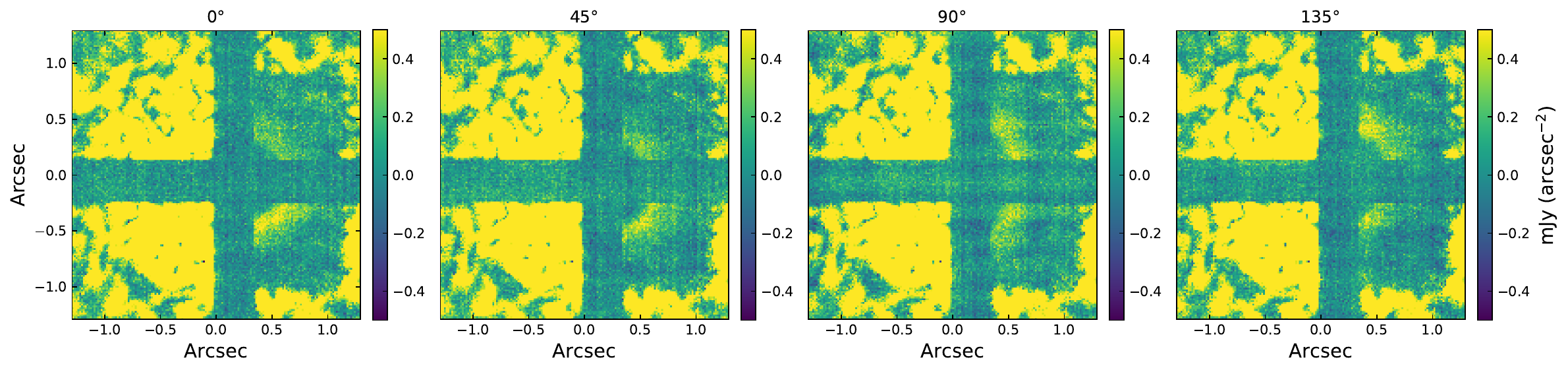}
   \caption{PSF-subtracted simulated images of Model A at four different roll angles: $0^\circ$, $45^\circ$, $90^\circ$, and $135^\circ$ (from left to right). Each panel shows the region within CPI-C's dark zones, which shift relative to the sky as the spacecraft rotates. The disk appears at different locations in the field of view, highlighting the importance of roll angle diversity for achieving complete coverage of the inner debris structure.}
   \label{fig:rotated}
\end{figure}

Due to the specialized design of CPI-C's dark zones, which cover only two squared regions in a single exposure, it is not possible to image the entire inner belt of $\epsilon$~Eri in one observation. To recover the complete morphology of the inner debris structure, we simulate multiple observations at different roll angles. Specifically, we assume eight roll angles spaced at $45^\circ$ intervals, with each observation consisting of 20 frames. Figure \ref{fig:rotated} presents four representative examples with roll angles of $0^\circ$, $45^\circ$, $90^\circ$, and $135^\circ$. In each case, different portions of the debris disk are revealed within the dark zones, demonstrating the necessity of roll diversity to fully map the inner belt.

\subsection{Searching for the warm dust}
\begin{figure}
   \centering
  \includegraphics[width=15cm, angle=0]{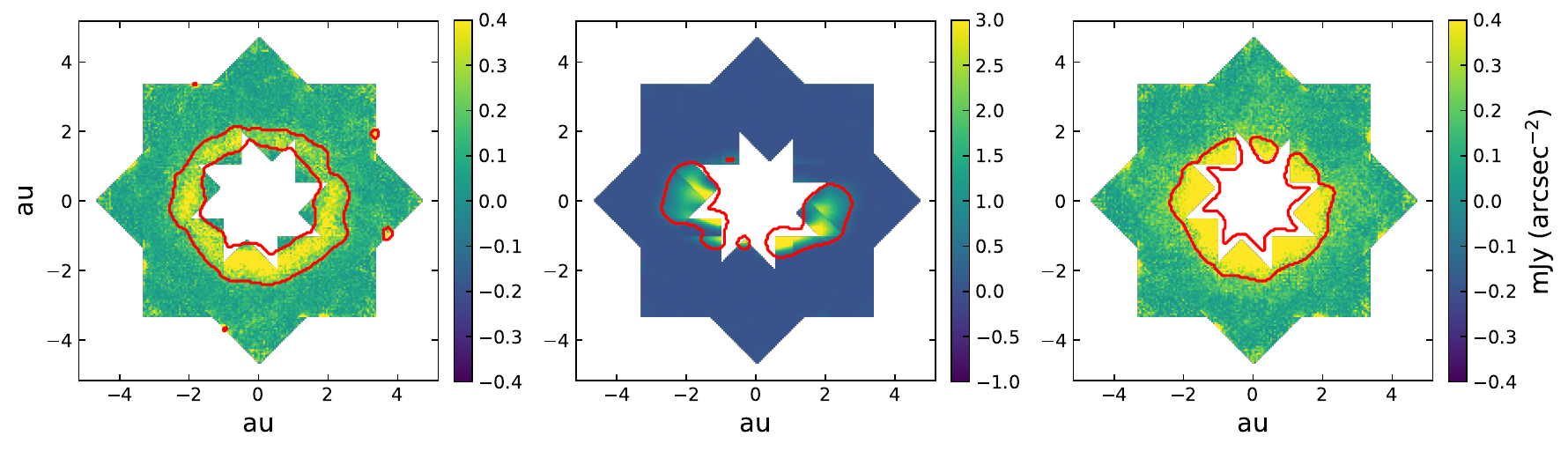}
   \caption{Stitched and smoothed images of the debris disk based on all eight roll-angle observations for three disk models: Model A (left), Model B (middle), and Model C (right). In each panel, individual dark-zone cutouts from different roll angles are re-rotated to a common sky orientation and combined to reconstruct the full inner belt. The red contours delineate the detected disk regions, used for morphological fitting. Model~A reveals the outer edges of a well-defined ring; Model~B exhibits a brighter, tilted structure due to strong forward scattering at high inclination; Model~C displays a more continuous, centrally peaked distribution with gradually decreasing brightness.}
   \label{fig:Stitched}
\end{figure}

To reconstruct a complete view of the inner debris belt, we extract and re-rotate the dark-zone regions from images with different roll-angle. These rotated segments are then stitched together to form a unified image for each disk model, as illustrated in Figure \ref{fig:Stitched}. The resulting mosaics represent the full-view synthetic images of the innermost disk, derived from eight roll angles. To enhance the visibility of morphological features, each stitched image is smoothed with a 3$\sigma$ Gaussian filter. Red contours outline the identified disk region based on brightness thresholds.

Assuming that the underlying disk structure is intrinsically ring-like, and thus appears as a projected elliptical ring, we estimate the inclination and radius of the disk for each model. In the left panel of Figure \ref{fig:Stitched}, corresponding to Model~A, the disk appears nearly symmetric, with a position angle close to $0^\circ$, consistent with the value adopted in the \texttt{MCFOST} simulation. The outer edge of the belt extends to approximately $\pm2.6$~au along the x-axis and $\pm2.3$~au along the y-axis, yielding a best-fit inclination of $28.4^\circ$, which is reasonably close to the input value of $34^\circ$ (see Table\ref{tab:mcfost}). The inner edge, however, lies near the observational limit of the CPI-C dark zone, with a maximum visible radius of only $\sim$ 1.7 au.

The middle panel displays the high-inclination model (Model~B, $i=70^\circ$). Owing to forward scattering, the observed disk flux is significantly brighter, consistent with predictions by \citet{Wolff2023}. The projected disk appears more inclined, with a derived position angle of $-6^\circ$, in good agreement with the input value of $-4^\circ$ \citep{Booth2017}. The extent along the x-axis is about $\pm2.7$ au, while coverage along the y-axis remain incomplete due to limited dark zone area, resulting in an estimated inclination of $63^\circ$, which is still reasonably close to the simulated value.

The right panel presents the results for the continuous disk model (Model~C). The spatial coverage is slightly smaller than that of Model A, due to the lower surface brightness in the edge of the disk. As shown in Table \ref{tab:mcfost}, the simulated disk radius is 3 au, whereas our derived radial extent reaches only $\pm2.3$ au. This underestimation is likely due to the decreased scattering efficiency at larger stellocentric distances. Nevertheless, the derived inclination of $29.6^\circ$ remains in good agreement with the input geometry.

\subsection{Searching for the cold planet}
As discussed in the previous section, no significant planetary signal is detected in the simulated images. This occurs because the planet's flux contrast lies near the detection limit of CPI-C and is overwhelmed by residual stellar speckles. To address this limitation, we consider imaging in polarized light, which has been demonstrated as an effective strategy to enhance planet detection \citep{Stam2004, Bailey2018}.

Reflected light from an exoplanet can exhibit a significant degree of linear polarization, particularly near quadrature, whereas stellar speckles remain largely unpolarized. Consequently, polarization differential imaging (PDI) offers a path to suppress the stellar background and enhance the planet's visibility. {CPIC is equipped with two orthogonal polarimeters of $0^\circ$ and $90^\circ$.} We simulate the polarized imaging of $\epsilon$ Eri b using the same optical configuration, assuming that the planet reflects light with a moderate degree of polarization based on its phase angle and atmospheric properties.

We adopt a simplified model of a gas-giant to estimate the polarization signal of $\epsilon$ Eri b, following the atmospheric configuration described in \citet{Tschudi2024}. The model comprises a Rayleigh scattering layer situated above a reflective cloud deck, capturing the primary features of polarized reflected light in a hydrogen-rich atmosphere.

\begin{figure*}
  \centering
\includegraphics[width=7cm, angle=0]{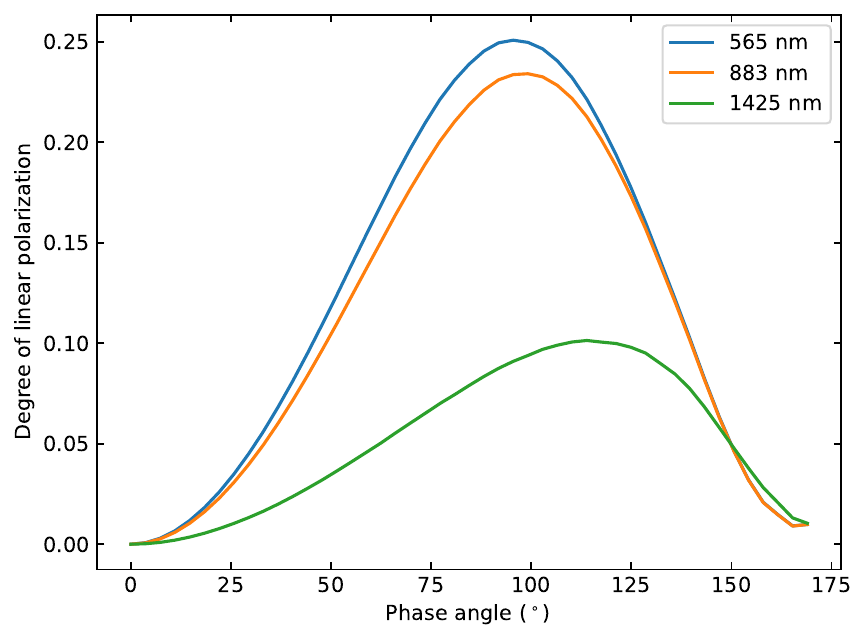}
\includegraphics[width=7cm, angle=0]{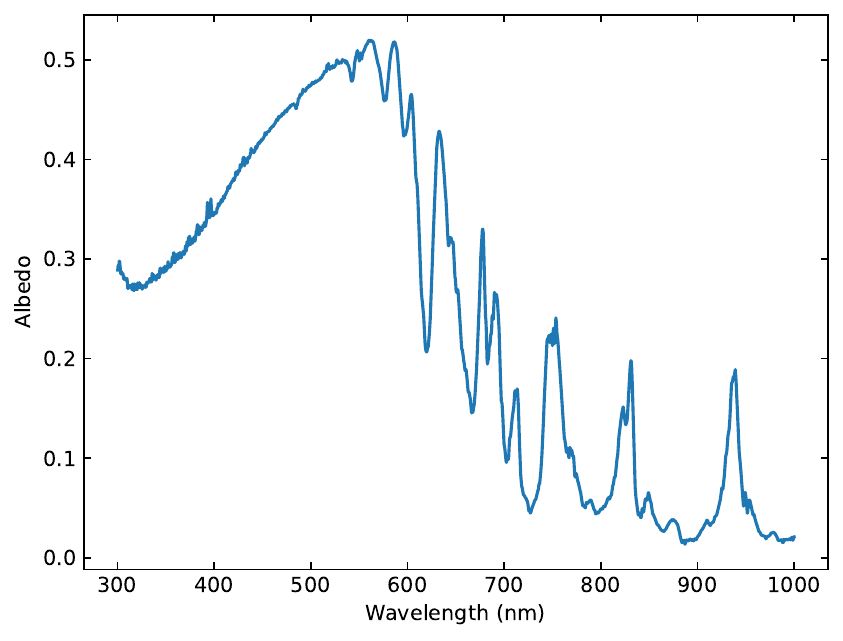}
   \caption{\textit{Left panel}: Degree of linear polarization as a function of planetary phase angle for three wavelength bands. The polarization reaches a maximum near $90^\circ$ phase angle, typical of Rayleigh-dominated atmospheres. \textit{Right panel}: Wavelength-dependent geometric albedo of the simulated gas giant planet model.}
   \label{fig:planet}
\end{figure*}

To compute the wavelength- and phase-dependent polarization properties, we employ the \texttt{PyMieDAP} package \citep{Rossi2018}, which solves the vector radiative transfer problem including multiple scattering. The left panel of Figure \ref{fig:planet} depicts the degree of linear polarization of $\epsilon$ Eri b as a function of phase angle for three representative wavelength bands. As expected, the degree of polarization peaks near a phase angle of $90^\circ$. The maximum polarization degree is approximately of 0.25, consists with the results by \citet{Tschudi2024}. Additionally, we estimate the albedo using a hybrid planet model \citep{Lacy2019}, assuming $\epsilon$ Eri b as a Jupiter analog.  The right panel of Figure \ref{fig:planet} shows the wavelength-dependent geometric albedo of the planet, peaking around 500 to 600 nm.

To assess the feasibility of detecting $\epsilon$ Eri b via polarization imaging, we simulate polarized observations using the \texttt{HCIPy} package \citep{Por2018}, incorporating the instrumental characteristics of CPI-C. In the simulation, we assume a 1\% instrumental polarization leakage. The polarized intensity of the planet is calculated as the product of the total reflected light contrast ($C_p$, as given in Equation~\ref{eq:contrast}) and the degree of linear polarization derived in the previous section.

\begin{figure}
   \centering
  \includegraphics[width=8cm, angle=0]{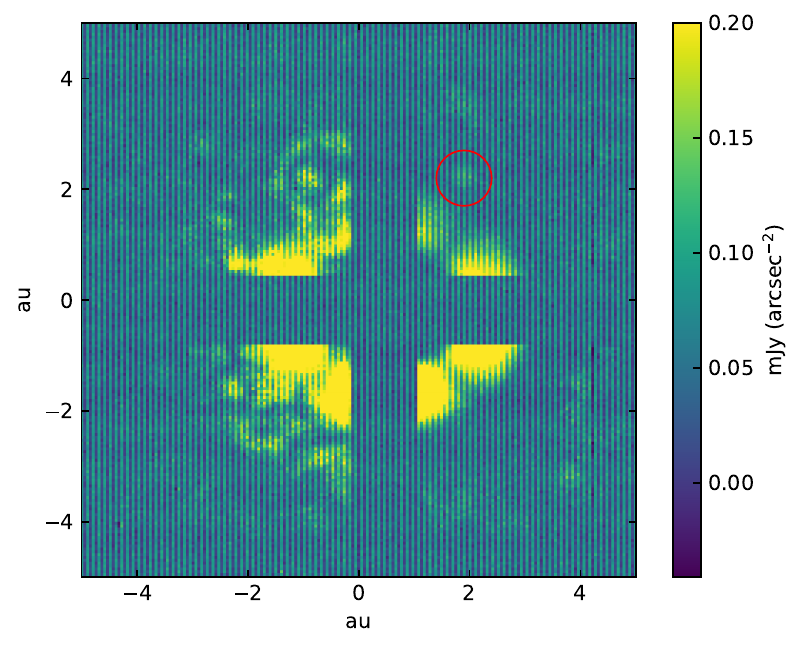}
   \caption{Simulated polarized intensity image of $\epsilon$ Eri observed with CPI-C in F565 band, based on the atmospheric polarization and hybrid albedo model. The planet is marked by the red circle.}
   \label{fig:polar_img}
\end{figure}

We then apply the same data reduction procedures described earlier, including dark and bias subtraction, cosmic ray removal, and PSF subtraction, to produce the final polarized image. Figure~\ref{fig:polar_img} shows the resulting image in a representative bandpass. The location of $\epsilon$ Eri b is indicated by a red circle, corresponding to a faint brightness enhancement due to the polarized planet signal.

Despite the enhanced contrast relative to unpolarized speckles, owing to the near-zero polarization of stellar speckles, the planetary signal remains challenging to detect due to a uniform instrumental bias pattern and low photon flux. We estimate that an exposure time of at least $\sim$ 300 seconds is required for a robust detection, which could be significantly affected by cosmic ray contamination in space-based detectors.

These results indicate that, although polarization imaging offers a promising pathway to enhancing planet detectability-especially in speckle-limited regimes-the polarimetric performance of CPI-C, particularly regarding bias calibration and cosmic ray mitigation, requires further detailed investigation and optimization. Techniques such as principal component analysis (PCA) \citep{Amara2012} and K-Stacker \citep{Nowak2018, Le2020} will assist in the data reduction process for CPI-C.

\section{Conclusions and Discussion}
\label{sect:Discussion}

In this study, we perform detailed simulations of high-contrast direct imaging of the $\epsilon$~Eridani system with CPI-C. By incorporating realistic models of the innermost debris belt and the long-period giant planet $\epsilon$~~Eri~~b, we evaluate CPI-C's capability to resolve complex circumstellar structures at optical wavelengths.

For the debris disk component, we model the ideal scattered-light intensity of the innermost belt with \texttt{MCFOST} with various parameters. Simulated observations are subsequently generated with \texttt{CPISM}. All three disk models are clearly resolved with CPI-C, and the disk extent and inclination are accurately recovered after post-processing. Our simulations show that strategical use of multiple roll angles in $45^\circ$ increments effectively rotates the dark zone to different sky positions, enabling complete coverage of the innermost belt.

For the planetary component, the detectability of $\epsilon$~~Eri~~b is highly dependent on orbital phase. We further assess the potential of polarimetric imaging to enhance detection, leveraging the polarized nature of planetary light compared to the nearly unpolarized stellar speckles. Using a simplified atmospheric model and \texttt{HCIPy} simulations that include instrumental polarization leakage, we find that polarization can improve the effective contrast of the planetary signal. Nevertheless, detection remains limited by systematic biases and low photon flux, necessitating long integrations that increase susceptibility to cosmic-ray contamination. These results highlight the need for further study of CPI-C's polarimetric performance, with emphasis on calibration accuracy and post-processing robustness.

{In addition to the potential for exoplanet detection, the polarimetric imaging is particularly powerful for debris disk studies, since scattered starlight from circumstellar dust is also strongly polarized. In CPI-C's observational design, linear polarization is measured only at $0^\circ$ and $90^\circ$. While this does not allow for a full reconstruction of the Stokes parameters, the differential signal between the two orientations provides a robust estimate of the polarized flux and enables us to trace the morphology of the debris disk.}

An optimized observing strategy for the $\epsilon$~Eri system would integrate angular diversity (roll maneuvers) and spectral diversity (multi-band imaging). Time-domain scheduling should prioritize orbital phases that maximize the phase angle-when reflected light is brightest-using prior knowledge of the planetary orbit \citep{Morgan2021, Bao2025}. $\epsilon$~Eri is also one of the prime targets for the Closeby Habitable Exoplanet Survey (CHES) mission \citep{Ji2022, Ji2024}, which aims to achieve sub-microarcsecond astrometric precision for nearby planetary systems \citep{Bao2024, Tan2024, HuangXM2025}. For $\epsilon$~Eri, the combination of CHES astrometry and CPI-C imaging would enable a comprehensive characterization of the giant planet $\epsilon$~Eri~b, serving as a valuable testbed for planet-disk interactions. Longer exposures, combined with multiple roll angles, will be essential for robust planet detection, while multi-epoch imaging of the debris disk will help distinguish static structures from features induced by planetary perturbations.

By comparison, the Roman Space Telescope will offer a coronagraph instrument (CGI) with similar contrast capabilities \citep{Poberezhskiy2020}. Roman's CGI is optimized for shorter wavelengths (500 $\sim$ 850~nm) and offers stable space-based wavefront control, likely achieving superior raw contrast. By contrast, CPI-C is tailored for multi-roll observations and flexible scheduling, providing advantages for characterizing highly inclined disks and disentangling planetary signals in complex circumstellar environments. Together, Roman and CPI-C are expected to yield complementary insights into the architectures of nearby planetary systems, particularly for dynamically rich targets such as $\epsilon$~Eri \citep{Anche2023}.

Overall, our results underscore CPI-C's unique potential to resolve the debris disk structures and giant planet around $\epsilon$~Eri. It is well-suited to test key dynamical predictions, including the coplanarity of debris belts and planetary orbits, and to probe the coevolution of planets and disks in nearby solar-type systems such as $\epsilon$~Eri. Future improvements in polarimetric calibration, PSF control, and long-exposure stability will further enhance CPI-C's scientific yields and guide the development of next-generation exoplanet imaging missions.

\normalem
\begin{acknowledgements}

   We thank the reviewer for their insightful comments and suggestions that improved the quality of the manuscript.
   This work is financially supported by the National Natural Science Foundation of China (grant Nos. 12033010 and 11773081), the China Manned Space Project (CMS-CSST-2025-A16, CMS-CSST-2021-B08,CMS-CSST-2021-B12), and the Foundation of Minor Planets of the Purple Mountain Observatory.

\end{acknowledgements}

\bibliographystyle{raa}
\bibliography{ms}

\end{document}